\documentclass[twocolumn,showpacs,preprintnumbers,amsmath,amssymb]{revtex4}
\usepackage{graphicx}% Include figure files
\usepackage{dcolumn}% Align table columns on decimal point
\usepackage{bm}% bold math

\begin{document}

\title{Deterministic scale-free networks created in a recursive manner}

\author{Zhongzhi Zhang}
\email{xinjizzz@sina.com}
\author{Lili Rong}
\email{llrong@dlut.edu.cn}

\affiliation{ Institute of Systems Engineering, Dalian University of
Technology,
2 Ling Gong Rd., Dalian 116023, Liaoning, China}%

\date{\today}

\begin{abstract}
In a recursive way and by including a parameter, we introduce a
family of deterministic scale-free networks. The resulting networks
exhibit small-world effects. We calculate the exact results for the
degree exponent, the clustering coefficient and the diameter. The
major points of our results indicate: the degree exponent can be
adjusted; the clustering coefficient of each individual vertex is
inversely proportional to its degree and the average clustering
coefficient of all vertices approaches to a nonzero value in the
infinite network order; and the diameter grows logarithmically with
the number of network vertices.
\end{abstract}

% 64.60.Ak, 02.50.Cw, 05.50.+q

%89.20.Hh World Wide Web, Internet
%89.75.Da Systems obeying scaling laws
%89.75.Fb Structures and organization in complex systems
%89.75.Hc Networks and genealogical trees
%89.75.-k Complex systems
%05.10.-a Computational methods in statistical physics and nonlinear
%                dynamics

\maketitle

%%%%%%%%%%%%%%%%%%%%%%%%%%%%%%%%%%%%%%%%%%%%%%%%%%%%%%%%%%%%%%%%%%%%

\section{Introduction}

Since the two seminal papers by Watts and Strogatz on small-world
networks \cite{WaSt98} and Barab\'asi and Albert on scale-free
networks \cite{BaAl99}, we have witnessed a considerable efforts
devoted to characterization and understanding of complex networks,
which describe many systems in nature and society. In the past few
years, there has been a substantial amount of interest in network
structure and function from a wide circle of researchers
\cite{St01,AlBa02,DoMe02,Ne03,SaVe04}. One particular question that
has attracted an exceptional amount of attention concerns the
structure of networks that are evolving over time. A number of
network models have been proposed, which convincingly reproduce some
or all features of real-life systems
\cite{St01,AlBa02,DoMe02,Ne03,SaVe04}.

Most of the precious models are stochastic
\cite{St01,AlBa02,DoMe02,Ne03,SaVe04}. However, because of their
advantages, deterministic networks have also received much
attention~\cite{BaRaVi01,IgYa05,DoGoMe02,JuKiKa02,CoFeRa04,RaSoMoOlBa02,RaBa03,No03,NaUeKaAk05,AnHeAnSi05,DoMa05,ZhCoFeRo05,CoOzPe00,CoSa02,ZhRoGo05,ZhWaHuCh04}.
First, the method of generating deterministic networks makes it
easier to gain a visual understanding of how networks are shaped,
and how do different vertices relate to each other; moreover,
deterministic networks allow to compute analytically their
properties: degree distribution, clustering coefficient, average
path length, diameter, betweenness, modularity and adjacency matrix
whose eigenvalue spectrum characterizes the
topology~\cite{BaRaVi01,IgYa05,DoGoMe02,JuKiKa02,CoFeRa04,RaSoMoOlBa02,RaBa03,No03,NaUeKaAk05,AnHeAnSi05,DoMa05,ZhCoFeRo05,CoOzPe00,CoSa02,ZhRoGo05,ZhWaHuCh04}.

The first model for deterministic scare-free networks was proposed
by Barab\'asi \emph{et al.} in Ref.~\cite{BaRaVi01} and was
intensively studied in Ref.~\cite{IgYa05}. Another elegant model,
called pseudofractal scale-free web (PSW)~\cite{DoGoMe02}, was
introduced by Dorogovtsev and Mendes, and was extended by Comellas
\emph{et al.} in Ref.~\cite{CoFeRa04}. Based on a similar idea of
PSW, Jung \emph{et al.} presented a class of recursive
trees~\cite{JuKiKa02}. Additionally, in order to discuss modularity,
Ravasz \emph{et al.} proposed a hierarchical network
model~\cite{RaSoMoOlBa02,RaBa03}, the exact scaling properties and
extensive study of which were reported in Refs.~\cite{No03}
and~\cite{NaUeKaAk05}, respectively. Recently, In relation to the
problem of Apollonian space-filing packing, Andrade \emph{et al.}
introduced Apollonian networks~\cite{AnHeAnSi05} which were also
proposed by Doye and Massen in Ref.~\cite{DoMa05} and have been
intensively investigated \cite{ZhCoFeRo05,ZhYaWa05,ZhRoCo05,ZhCo05}.
Except for above models, deterministic networks can be created by
various techniques: modification of some regular
graphs~\cite{CoOzPe00}, addition and product of
graphs~\cite{CoSa02}, edge iterations~\cite{ZhRoGo05} and other
mathematical methods as in Ref.~\cite{ZhWaHuCh04}.

As mentioned by Barab\'asi \emph{et al.}, it would be of major
theoretical interest to construct deterministic models that lead to
scale-free networks~\cite{BaRaVi01}. In this paper, we do an
extensive study on pseudofractal scale-free web~\cite{DoGoMe02} and
the recursive graphs (RG)~\cite{CoFeRa04}. In a simple recursive way
we propose a general model for PSW and RG by including a parameter,
with PSW and RG as particular cases of the present model. The
deterministic construction of our model enables one to obtain the
analytic solutions for some main structure properties: degree
distribution, clustering coefficient and diameter. By adjusting the
parameter, we can obtain a variety of scale-free networks.

%%%%%%%%%%%%%%%%%%%%%%%%%%%%%%%%%%%%%%%%%%%%%%%%%%%%
%%The iterative algorithm for high-dimensional Apollonian networks
%%%%%%%%%%%%%%%%%%%%%%%%%%%%%%%%%%%%%%%%%%%%%%%%%%%
\section{The network construction}
%
%\emph{Algorithm for the networks}.
Before introducing our networks we give the following definitions on
a graph. The term \emph{size} refers to the number of edges in a
graph. The number of vertices in a graph is called its \emph{order}.
When two vertices of a graph are connected by an edge, these
vertices are said to be \emph{adjacent}, and the edge is said to
join them. A \emph{complete graph} is a graph in which all vertices
are adjacent to one another. Thus, in a complete graph, every
possible edge is present. The complete graph with $q$ vertices is
denoted as $K_q$ (also referred in the literature as
$q$-\emph{clique}; see \cite{We01}). Two graphs are
\emph{isomorphic} when the vertices of one can be relabeled to match
the vertices of the other in a way that preserves adjacency. So all
$q$-cliques are isomorphic to one another.

\begin{figure}
\begin{center}
\includegraphics[width=12cm]{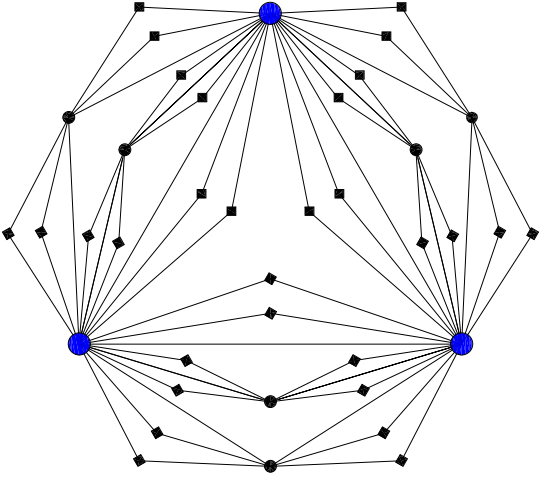}
\caption{Scheme of the growth of the deterministic network for the
case of $q=2$ and $m=2$. Only the first three steps are shown. }
\label{recursive}
\end{center}
\end{figure}

The network is constructed in a recursive way. We denote the network
after $t$ steps by $R(q,t)$, $q\geq 2, t\geq 0$ (see Fig.
\ref{recursive}). Then the network at step $t$ is constructed as
follows: For $t=0$, $R(q,0)$ is a complete graph $K_{q+1}$ (or
$(q+1)$-clique) consist of $q+1$ $q$-cliques), and $R(q,0)$ has
$q+1$ vertices and $q(q+1)/2$ edges. For $t\geq 1$, $R(q,t)$ is
obtained from $R(q,t-1)$ by adding $m$ new vertices for each of its
existing subgraphs isomorphic to a $q$-clique, and each new vertex
is connected to all the vertices of this subgraph. In the special
case $m=1$, it is reduced to the network described in Ref.
\cite{CoFeRa04} which is a generalization of pseudofractal
scale-free web~\cite{DoGoMe02}.

%%%%%%%%%%%%%%%%%%%%%%%%%%%%%%%%%%%%%%%%%%%%%%%%%%%%%%%%%%
% Figure  1
%%%%%%%%%%%%%%%%%%%%%%%%%%%%%%%%%%%%%%%%%%%%%%%%%%%%%%%%%%
%%%%%%%%%%%%%%%%%%%%%%%%%%%%%%%%%%%%%%%%%%%%%%%%%%%%%%%%%%

Let $n_v(t)$ and $n_e(t)$ be the number of vertices and edges
created at step $t$, respectively. Denote $K_{q,t}$ as the total
number of $q$-cliques in the whole network at step $t$. Note that
the addition of each new vertex leads to $q$ new $q$-cliques and $q$
new edges. By construction, we have $n_e(t)=qn_v(t)$,
$n_v(t)=mK_{q,t-1}$ and $K_{q,t}=K_{q,t-1}+qn_v(t)$. Thus one can
easily obtain $K_{q,t}=(mq+1)K_{q,t-1}=(q+1)(mq+1)^{t}$ ($t\geq 0$),
$n_v(t)=m(q+1)(mq+1)^{t-1}$ ($t>0$) and $n_e(t)=mq(q+1)(mq+1)^{t-1}$
($t>0$). So the number of network vertices increases exponentially
with time. From above results, we can easily compute the size and
order of the networks. The total number of vertices $N_t$ and edges
$|E|_t$ present at step $t$ is
\begin{eqnarray}\label{Nt}
N_t=\sum_{t_i=0}^{t}n_v(t_i)=\frac{(q+1)[(mq+1)^{t}+q-1]}{q}
\end{eqnarray}
and
\begin{eqnarray}\label{Et}
|E|_t
=\sum_{t_i=0}^{t}n_e(t_i)=(q+1)(mq+1)^{t}+\frac{(q+1)(q-2)}{2},
\end{eqnarray}
respectively. So for large $t$, The average degree $\overline{k}_t=
\frac{2|E|_t}{N_t}$ is approximately $2q$.

\section{Characteristics of the networks}

%Below we will find that the dimension $d$ is a tunable parameter
%controlling all the relevant characteristics of the $d$-dimensional
%Apollonian network.
\subsection{Degree distribution}
%\emph{Degree distribution}.
%\emph{Degree distribution}.
When a new vertex $i$ is added to the graph at step $t_i$, it has
degree $q$ and forms $q$ new $q$-cliques. Let $n_q(i,t)$ be the
total number of $q$-cliques at step $t$ that will created new
vertices connected to the vertex $i$ at step $t+1$. So at step
$t_i$, $n_q(i,t_i)=q$. By construction, we can see that in the
subsequent steps each new neighbor of $i$ generated $q-1$ new
$q$-cliques with $i$ as one vertex of them. Let $k_i(t)$ be the
degree of $i$ at step $t$. It is not difficult to find following
relations for $t>t_i+1$:
\begin{equation}
\Delta k_i(t)=k_i(t)-k_i(t-1)=mn_q(i,t-1)
\end{equation}
and
\begin{equation}
n_q(i,t)=n_q(i,t-1)+(q-1)\Delta k_i(t).
\end{equation}
From the above two equations, we can derive
$n_q(i,t_i)=[m(q-1)+1]n_q(i,t_i-1)$. Considering $n_q(i,t_i)=q$, we
obtain $n_q(i,t)=q[m(q-1)+1]^{t-t_i}$ and $\Delta
k_i(t)=mq[m(q-1)+1]^{t-t_i-1}$. Then the degree $ k_i(t)$ of vertex
$i$ at time $t$ is
\begin{eqnarray}\label{Ki}
k_i(t)&=&k_i(t_i)+\sum_{t_h=t_i+1}^{t}{\Delta k_i(t_h)}\nonumber\\
&=&\frac{q[m(q-1)+1]^{t-t_i}+q^{2}-2q}{q-1}.
\end{eqnarray}
Since the degree of each vertex has been obtained explicitly as in
Eq.~(\ref{Ki}), we can get the degree distribution via its
cumulative distribution, i.e. $P_{cum}(k) \equiv \sum_{k^\prime \geq
k} N(k^\prime,t)/N_t \sim k^{1-\gamma}$, where $N(k^\prime,t)$
denotes the number of vertices with degree $k^\prime$. The analytic
computation details are given as follows. For a degree $k$
\begin{equation}
k=\frac{q[m(q-1)+1]^{t-j}+q^{2}-2q}{q-1},
\end{equation}
there are  $n_v(j)=m(q+1)(qm+1)^{t-1}$ vertices with this exact
degree, all of which were born at step $j$. All vertices with birth
time at $j$ or earlier have this and a higher degree. So we have
\begin{equation}
\sum_{k' \geq k} N(k',t)=\sum_{s=0}^{j}n_v(s)=\frac{(q+1)[(mq+1)^{j}+q-1]}{q}. \nonumber\\
\end{equation}
As the total number of vertices at step $t$ is given in
Eq.~(\ref{Nt}) we have
\begin{equation}
\left[\frac{q[m(q-1)+1]^{t-j}+q^{2}-2q}{q-1}\right]^{1-\gamma}\nonumber\\
=\frac{\frac{(q+1)[(mq+1)^{j}+q-1]}{q}}{\frac{(q+1)[(mq+1)^{t}+q-1]}{q}}.\nonumber\\
\end{equation}
Therefore, for large $t$ we obtain
\begin{equation}
\left [[m(q-1)+1]^{t-j}\right ]^{1-\gamma}=(mq+1)^{j-t}
\end{equation}
and
\begin{equation}\label{gamma}
\gamma \approx 1+\frac{\ln (mq+1)}{\ln[m(q-1)+1]}.
\end{equation}
For the particular case of $m=1$, Eq. (\ref{gamma}) recovers the
results previously obtained in Ref. \cite{CoFeRa04}.

\subsection{Clustering coefficient}

%\emph{Clustering coefficient}.
The clustering coefficient~\cite{WaSt98} $ C_i $ of vertex $i$ is
defined as the ratio between the number of edges $e_i $ that
actually exist among the $k_i $ neighbors of vertex $i$ and its
maximum possible value, $ k_i( k_i -1)/2 $, i.e., $ C_i =2e_i /k_i(
k_i -1)$. The clustering coefficient of the whole network is the
average of $C_i's$ over all vertices in the graph.

For our networks, the analytical expressions for clustering
coefficient $C(k)$ of the individual vertex with degree $k$ can be
derived exactly. When a vertex is created it is connected to all the
vertices of a $q$-clique whose vertices are completely
interconnected. Its degree and clustering coefficient are $q$ and 1,
respectively. In the following steps, if its degree increases one by
a newly created vertex connecting to it, then there must be $q-1$
existing neighbors of it attaching to the new vertex at the same
time. Thus for a vertex of degree $k$, we have
\begin{equation}\label{Ck}
C(k)= {{{q(q-1)\over 2}+ (q-1)(k-q)} \over {k(k-1)\over 2}}=
\frac{2(q-1)(k-\frac{q}{2})}{k(k-1)},
\end{equation}
which depends on degree $k$ and $q$. For $k \gg q$, the $C(k)$ is
inversely proportional to degree. The scaling $C(k)\sim k^{-1}$ has
been found for some network
models~\cite{DoGoMe02,CoFeRa04,RaBa03,No03,AnHeAnSi05,DoMa05,ZhCoFeRo05,ZhYaWa05,ZhRoCo05},
and has also observed in several real-life networks~\cite{RaBa03}.

Using Eq. (\ref{Ck}), we can obtain the clustering $\overline{C}_t$
of the networks at step $t$:
%%%%%%%%%
%\begin{eqnarray*}
%\lefteqn{\overline{S}_t =\frac{2(q+1)(q-1)(\Delta_t-\frac{q}{2})}{\Delta_t(\Delta_t-1)}}\\
%     &+& \sum_{i=1}^{t-1} \frac{2(q+1)^{t-i}(q-1)(\Delta_i-\frac{q}{2})}{\Delta_i(\Delta_i-1)}
%\end{eqnarray*}
\begin{equation}\label{AC}
\overline{C}_t=
    \frac{1}{N_{t}}\sum_{r=0}^{t}
    \frac{2(q-1)(D_r-\frac{q}{2})n_v(r)}{D_r(D_r-1)},
\end{equation}
where the sum is the total of clustering coefficient for all
vertices and $D_r=\frac{q[m(q-1)+1]^{t-r}+q^{2}-2q}{q-1}$ shown by
Eq. (\ref{Ki}) is the degree of the vertices created at step $r$.

\begin{figure}
\begin{center}
\includegraphics[width=8cm]{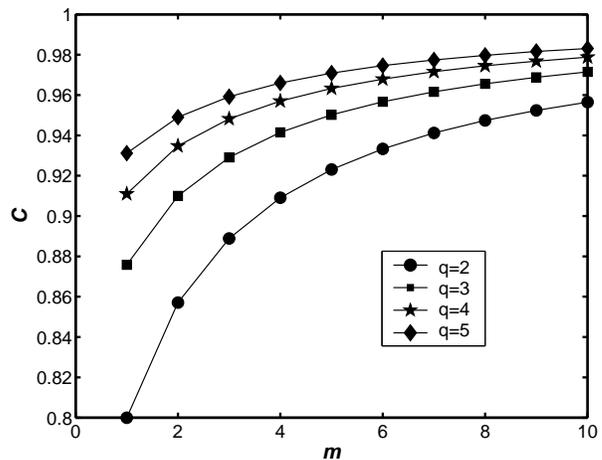}
\caption{The dependence relation of $C$ on $q$ and $m$.} \label{cc}
\end{center}
\end{figure}

It can be easily proved that for arbitrary fixed $m$,
$\overline{C}_t$ increases with $q$, and that for arbitrary fixed
$q$, $\overline{C}_t$ increases with $m$. In the infinite network
order limit ($N_{t}\rightarrow \infty$), Eq. (\ref{AC}) converges to
a nonzero value $C$. When $q=2$, for $m=1$, 2, 3 and 4, $C$ equal to
0.8000, 0.8571 0.8889 and 0.9091, respectively. When $m=2$, for
$q=2$, 3, 4 and 5, $C$ are 0.8571, 0.9100, 0.9348 and 0.9490,
respectively. Therefore, the clustering coefficient of our networks
is very high. Moreover, similarly to the degree exponent $\gamma$,
clustering coefficient $C$ is determined by $q$ and $m$. Fig.
\ref{cc} shows the dependence of $C$ on $q$ and $m$.

%%%%%%%%%%%%%%%%%%%%%%%%%%%%%%%%%%%%%%%%%%%%%%%%%%%%%%%%%%
% Figure  2
%%%%%%%%%%%%%%%%%%%%%%%%%%%%%%%%%%%%%%%%%%%%%%%%%%%%%%%%%%
%%%%%%%%%%%%%%%%%%%%%%%%%%%%%%%%%%%%%%%%%%%%%%%%%%%%%%%%%%
%%%%%%%%%%%%%%%%%%%%%%%%%%%%%%%%%%%%%
\subsection{Diameter}
The diameter of a network characterizes the maximum communication
delay in the network and is defined as the maximum of shortest path
between all pairs of vertices. In what follows, the notations $
\lceil x \rceil$ and $\lfloor x \rfloor$ express the integers
obtained by rounding $x$ to the nearest integers towards infinity
and minus infinity, respectively. Now we compute the diameter of
$R(q,t)$, denoted $Diam(R(q,t))$ for $q\geq 3$ ($q=2$ is a
particular case that is treated separately):

{\em Step 0}.  The diameter is $1$.

{\em Steps 1 to $\lceil\frac{q+1}{2}\rceil-1$}.  In this case, the
diameter is 2, since any new vertex is by construction connected to
a $q$-clique forming a $(q+1)$-clique, and since any $(q+1)$-clique
during those steps contains at least $\lceil\frac{q+1}{2}\rceil$
($q$ even) or $\lceil\frac{q+1}{2}\rceil$+1 ($q$ odd) vertices from
the initial $(q+1)$-clique $R(q,0)$ obtained after step 0. Hence,
any two newly added vertices $u$ and $v$ will be connected
respectively to sets $S_u$ and $S_v$, with $S_u\subseteq V(R(q,0))$
and $S_v\subseteq V(R(q,0))$,  where $V(R(q,0))$ is the vertex set
of $R(q,0)$; however, since $\vert
S_u\vert\geq\lceil\frac{q+1}{2}\rceil$ ($q$ even) and $\vert
S_v\vert\geq\lceil\frac{q+1}{2}\rceil$+1 ($q$ odd), where $\vert
S\vert$ denotes the number of elements in set $S$, we conclude that
$S_u\cap S_v\neq{\O}$, and thus the diameter is 2.

{\em Steps $\lceil\frac{q+1}{2}\rceil$ to $q$}. In any of those
steps, some newly added vertices might not share a neighbor in the
original $(q+1)$-clique $R(q,0)$; however, any newly added vertex is
connected to at least one vertex of the initial $(q+1)$-clique
$R(q,0)$. Thus, the diameter is equal to 3.

{\em Further steps}. In order to simplify the analysis, we first
note that it is unnecessary to look at all the vertices in the graph
in order to find the diameter. In other words, some vertices added
at a given step can be ignored, because they do not increase the
diameter from the previous step. These vertices are those that
connect to vertices that already formed a $(q+1)$-clique in one or
several of the previous steps: indeed, for these vertices we know
that a similar construction has been done in previous steps, so we
can ignore them for the computation of the diameter. Let us call
``outer'' vertices the vertices which are connected to a $q$-clique
that did not exist previously. Clearly, at each step, the diameter
depends on the distances between outer vertices.

 Now, at any step
$t\geq q+1$, we note that an outer vertex cannot be connected with
two or more vertices that were created during the same step
$0<t'\leq t-1$. Moreover, by construction no two vertices that were
created during a given step are neighbors, thus they cannot be part
of the same $q$-clique. Therefore, for any step $t\geq q+1$, some
outer vertices are connected with vertices that appeared at pairwise
different steps. Thus, if $v_t$ denotes an outer vertex that was
created at step $t$, then $v_t$ is connected to vertices $v_i$s,
$1\leq i\leq t-1$, where all the $i$s are pairwise distinct. We
conclude that $v_t$ is necessarily connected to a vertex that was
created at a step $t_0\le t-q$. If we repeat this argument, then we
obtain an upper bound on the distance from $v_t$ to the initial
$(q+1)$-clique $R(q,0)$. Let $t=\alpha q+p$, where $1\leq p\leq q$.
Then, we see that $v_t$ is at distance at most $\alpha +1$ from a
vertex in $R(q,t)$. Hence any two
 vertices $v_t$ and $w_t$ in $R(q,0)$ lie at distance at most
$2(\alpha +1)+1$ ; however, depending on $p$, this distance can be
reduced by 1, since when $p\leq \lceil\frac{q+1}{2}\rceil-1$, we
know that two vertices created at step $p$ share at least a neighbor
in $R(q,0)$. Thus, when $1\leq p\leq \lceil\frac{q+1}{2}\rceil-1$,
$Diam(R(q,t))\leq 2(\alpha +1)$, while when
$\lceil\frac{q+1}{2}\rceil\leq p\leq q$, $Diam(R(q,t))\leq 2(\alpha
+1)+1$. One can see that these bounds can be reached by pairs of
outer vertices created at step $t$. More precisely, those two
vertices $v_t$ and $w_t$ share the property that they are connected
to $q$ vertices that appeared respectively at steps $t-1,t-2,\ldots
t-q$.

Based on the above arguments, one can easily see that for $t>q$, the
diameter increases by 2 every $q$ steps. More precisely, we have the
following result, for any $q\geq 3$ and $t\geq 1$ (when $t=0$, the
diameter is clearly equal to 1):
$$Diam(K(q,t))=2(\lfloor\frac{t-1}{q}\rfloor +1)+f(q,t),$$
where $f(q,t)=0$ if $t-\lfloor\frac{t-1}{q}\rfloor q\leq
\lceil\frac{q+1}{2}\rceil-1$, and 1 otherwise. When $t$ gets large,
$Diam(R(q,t))\sim \frac{2t}{q}$, while $N_t\sim (mq+1)^{t}$, thus
the diameter grows logarithmically with the number of vertices.

For $q=2$, the argument is similar: when $t=0$ (resp. $t=1$), the
diameter is equal to 1 (resp. 2). Now consider two outer vertices
created at step $t\geq 2$, say $v_t$ and $w_t$. Then $v_t$ is
connected to two vertices, and one of them must have been created
before or during step $t-2$. We repeat this argument, and we end up
with 2 cases: (1) $t=2m$ is even. Then, if we make $m$ ``jumps'',
from $v_t$ we arrive in $R(2,0)$, in which we can reach any $w_t$ by
using an edge of $R(2,0)$ and making $m$ jumps to $w_t$ in a similar
way. Thus $Diam(R(2,2m))\leq 2m+1$. (2) $t=2m+1$ is odd. In this
case we can stop after $m$ jumps at $R(2,1)$, for which we now that
the diameter is 2, and make $m$ jumps in a similar way to reach
$w_t$. Thus $Diam(R(2,2m+1))\leq 2m+2$. As previously, it easily
seen that the bound can be reached for some pairs of vertices.

Hence, formally, $Diam(R(2,t))=t+1$ for any $t\geq 0$. In this
particular case, the network order $N_t\sim (2m+1)^{t}$, thus
$Diam(R(2,t))$ also increases logarithmically with network order.

\section{Conclusion and discussion}

 %\emph{Conclusion and discussion.}
In summary, we have proposed and studied a network model constructed
in a recursive fashion. At each time step, each already existing
$q$-clique generates $m$ new vertices. The process results in a
serial of networks with two published papers
~\cite{DoGoMe02,CoFeRa04} as special cases of them. We have obtained
the analytical results for degree exponent, clustering coefficient
and diameter. The degree exponent and the clustering coefficient may
be adjusted to various values by tuning the parameter $m$.
Therefore, they may perform well in mimicking a variety of
scale-free networks in real world. Moreover, in a similar way, one
can easily consider other variations, for example, at each step not
all cliques of a network but only some of them generate vertices,
which also allow a rich structure and flexibility in the control of
degree exponent, clustering coefficient and other properties.
Details of the analytical solution for these variants will be
addressed elsewhere.
\smallskip
\smallskip

This research was supported by the National Natural Science
Foundation of China under Grant No. 70431001. %The authors are grateful to the anonymous referees
%for their valuable comments and suggestions.

\end{document}